\documentclass[onecolumn,sort&compress,numbers]{els-mrw} % For numbered references Style 

\usepackage{amsmath,amssymb,amsfonts,amsthm,makeidx,graphicx}
\usepackage{txfonts}
\usepackage{helvet}
\usepackage{graphicx} % Required for inserting images
\usepackage{subcaption}
\usepackage{slashed}
\usepackage{bbold}
\usepackage[hidelinks]{hyperref}
\usepackage[utf8]{inputenc}
\usepackage[T1]{fontenc}
\usepackage{lmodern}
\usepackage{xspace}
\usepackage{color}
\usepackage[dvipsnames]{xcolor}
\usepackage{setspace}
\usepackage{footmisc}
\usepackage{comment}

\usepackage{lineno}
\newcommand\NOARXIV[1]{}

\newcommand{\secref}[1]{Sec.\,\,\ref{#1}}  %Reference to a section
\newcommand{\subsecref}[1]{Subsec.\,\,\ref{#1}}  %Reference to a subsection
\newcommand{\figref}[1]{Fig.\,\,\ref{#1}}  %Reference to a figure
\newcommand{\tabref}[1]{Table\,\,\ref{#1}}  %Reference to a table
\begin{document}

%%%%%%%%%%%%%%%%%%%%%%%%%%%%%%%%%%%%%%%%%%%%%%%%%%%%%%%%%%%%%%%%
%% the following items are mandatory: 
%% - title
%% - author names
%% - affiliation details
%% - abstract
%% - keywords

%% Precise, concise, and informative description of the focus of this work. Avoid abbreviations and formulae in the title
\chapter{Monte Carlo Event Generators}\label{chapx}

\author[1]{J\"urgen Reuter}

\address[1]{\orgname{Deutsches Elektronen-Synchrotron DESY}, \orgaddress{Notkestr. 85, 22607 Hamburg,
Germany}}

\articletag{Chapter Article tagline: update of previous edition, reprint.}

\maketitle

%%%%%%%%%%%%%%%%%%%%%%%%%%%%%%%%%%%%%%%%%%%%%%%%%%%%%%%%%%%%%%%%
%% the following item is mandatory: 
%% 100-150 word summary of the chapter
\par \vspace{2mm}
\begin{abstract}[Abstract]
  Monte Carlo event generators are in a modern terminology the digital
twins of collider-based particle physics experiment. We give an
introduction into the application of MC generators for particle
physics, discuss their different components each simmulating a
different aspect of physics. The main part is the hard scattering
process, sampled over a commplicated phase space of kinematic
variables, followed by simulation of strong and electromagnetic
radiation in parton showers, and hadronization of the end products of
the shower into mesons and baryons. These then undergo several levels
of decays into those particles that are measured in the detectors --
mostly electrons, photons, pions and kaons. For each step, the main
techniques will be described and explained.

\end{abstract}

%% 5-10 words that embody the key topics in the chapter. What terms would someone put into a search engine if they were looking for a chapter like this?
\begin{keywords}
%       please enter 5 keywords as follows:
        Monte Carlo generators\sep Parton shower\sep Hadronization\sep
        phase space sampling \sep higher order corrections
\end{keywords}

\vspace*{\fill}

%%% \begin{center}
%%%   {\em Invited contribution to Encyclopedia of Particle Physics}
%%% \end{center}

%% \begin{flushleft}
%% May 2025
%% \end{flushleft}

%%%%%%%%%%%%%%%%%%%%%%%%%%%%%%%%%%%%%%%%%%%%%%%%%%%%%%%%%%%%%%%%
%% the following item is optonal: 
%% - Single figure visually illustrating the key topic/method/outcome described in the chapter
%\begin{figure}[h]
%       \centering
        %\includegraphics[width=7cm,height=4cm]{blankfig}
%       \caption{Optional: Single figure visually illustrating the key topic/method/outcome described in the chapter. 
%            Please add here some text explaining the pic...}
%\label{fig:titlepage}
%\end{figure}

%%%%%%%%%%%%%%%%%%%%%%%%%%%%%%%%%%%%%%%%%%%%%%%%%%%%%%%%%%%%%%%%
%% the following item is optional: 
%% - System of abbreviations/terms/symbols used in the specific field of study/community. List and define
%\begin{glossary}[Nomenclature]
%       \begin{tabular}{@{}lp{34pc}@{}}
%               LHC & Large Hadron Collider\\
%       \end{tabular}
%\end{glossary}

%%%%%%%%%%%%%%%%%%%%%%%%%%%%%%%%%%%%%%%%%%%%%%%%%%%%%%%%%%%%%%%%
%% the following item is mandatory: 
%% List of the key points and topics a reader can expect to learn from this chapter 
\section*{Objectives}
\begin{itemize}
        \item  Monte Carlo event generators are introduced as the main
          working horses of collider physics experiments in particle physics.
        \item Monte Carlo integration, sampling of multi-dimensional
          phase spaces and event generation are discussed.
        \item The generation of amplitudes as building blocks and
          their higher order extensions are introduced.
        \item The simulation of proton and lepton distribution
          functions is discussed.
        \item Parton showers and their matching to fixed order
          perturbation theory and hadronization are explained in detail.
        \item Some special topics, software development, performance
          and efficiency aspects are listed.
    \item 
    %Some text on the third topic of the chapter, i.e. what else the reader can learn from this article
\end{itemize}

\section{Introduction}\label{intro}
\begin{figure}[t]
  \centering
  \includegraphics[width=0.35\textwidth]{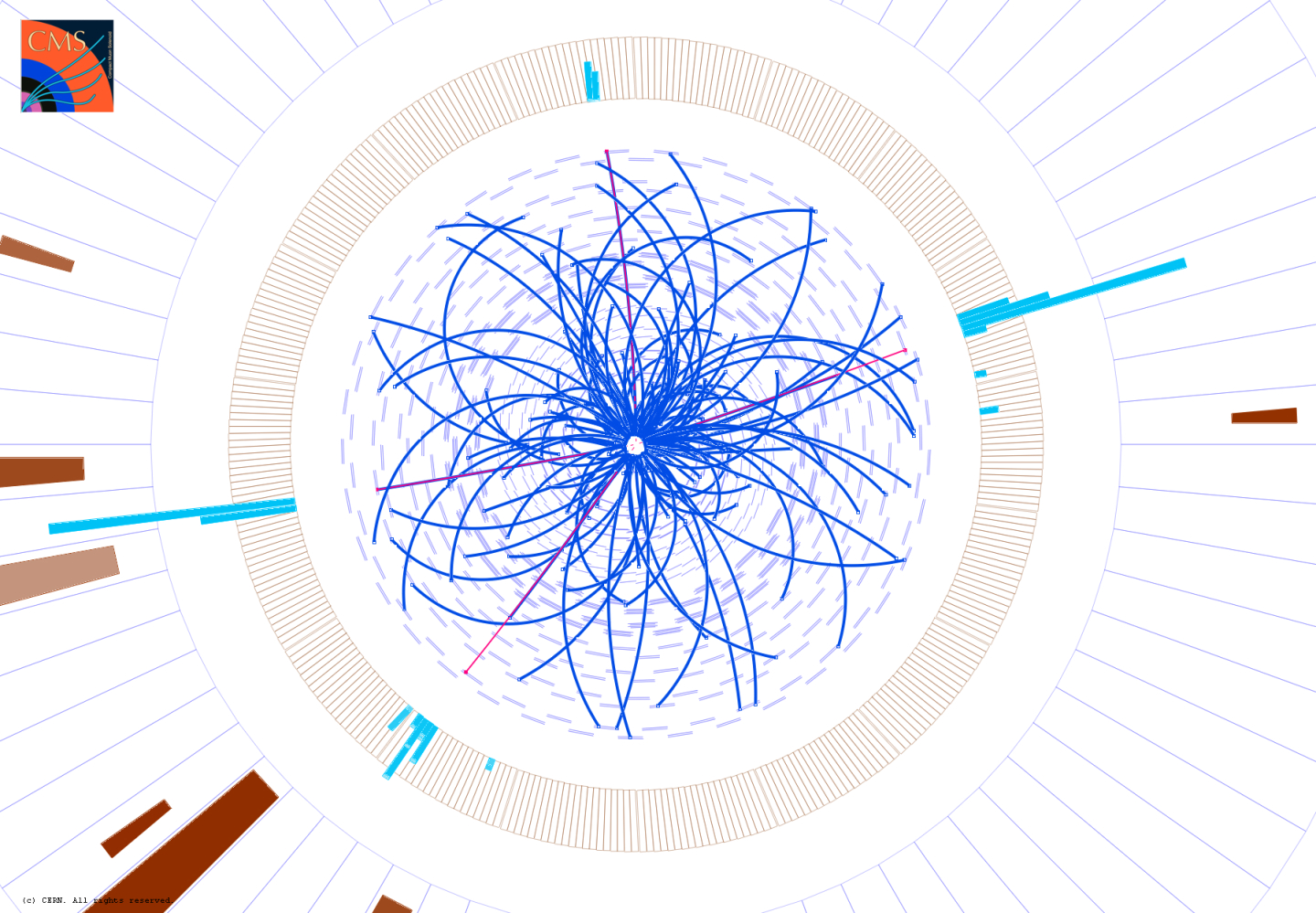}\,
  \includegraphics[width=0.55\textwidth]{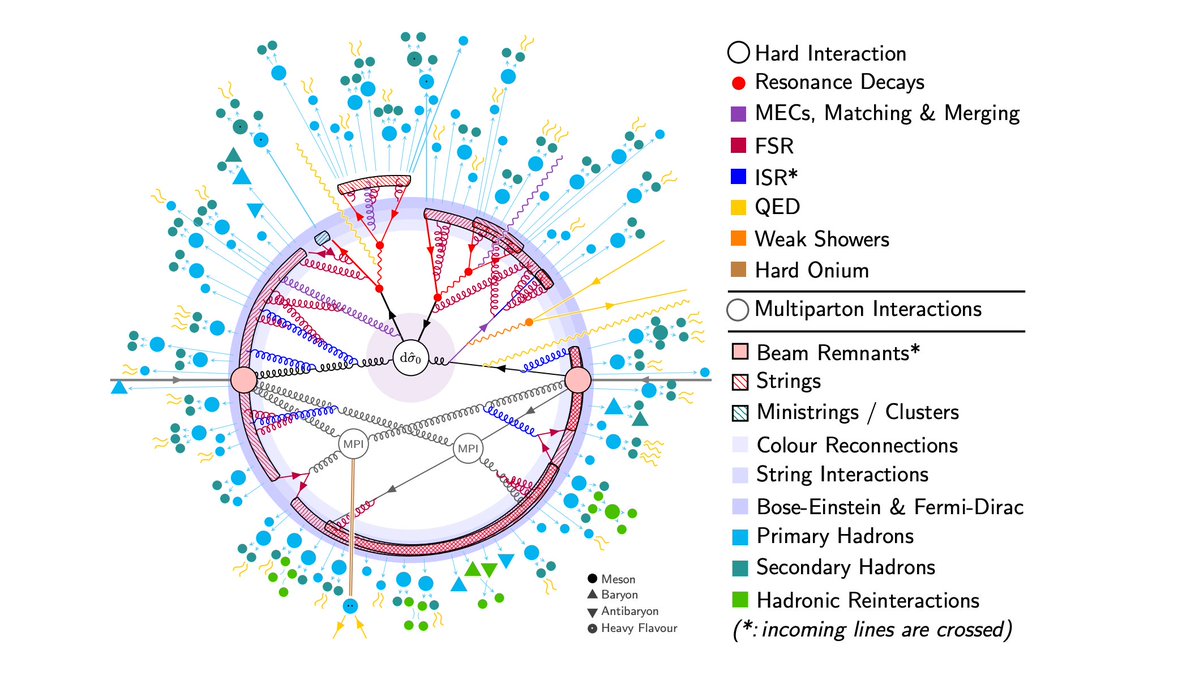}
  \caption{Particle physics event display (from the CMS detector at
  the Large Hadron Collider (LHC) (left), and different components of
  its digital twin, the MC generator simialtion (right).}
  \label{fig:events_mcgen}
\end{figure}

Monte Carlo (MC) event generators take all of our knowledge about the
fundamental interactions and matter in particle physics, quantum
chromodynamics (QCD) for strong interactions and electroweak (EW)
interactions (weak interactions and quantum electrodynamics (QED), and
encode them in software tools to simulate collider or fixed-target
experiments. There are big synergies with MC generators for air
showers in cosmic ray physics (in terms of physics) and detector
simulation (in terms of methods and also physics); both topics are
beyond the scope of this section. Fig.~\ref{fig:events_mcgen} shows an
event display of a particle physics (a Higgs candidate event from the
CMS detector at the Large Hadron Collider LHC) on the left, while the
right hand side shows the part simulated by a MC generator. All
components on the left are within the small white circle in the center
(and partially in the blue bent lines in the tracker volumes). The
sketch on the right of Fig.~\ref{fig:events_mcgen} has been designed
for hadron collisions, but can easily be transferred to
electron-positron, muon or electron-hadron collisions, respectively.

We will first discuss the rationale of Monte Carlo generators for
sampling highdimensional phase spaces and do weighted and unweighted
event generation in Sec.~\ref{sec:mcgen}. The core physics part is the
so called hard scattering process, e.g. $gg \to H \to \gamma\gamma$ 
for the discovery channel of the Higgs boson at the LHC in 2012. In
Sec.~\ref{sec:hardscat} we will discuss methods to generate code for
matrix elements at fixed perturbative order (tree-level and beyond) in
an effective manner. In addition, phase space integration over
processes with high multiplicities of final states with complicated
kinematical cuts will be explained. Furthermore, we will connect this
to methods to analytically or semi-numerically resum perturbative
corrections to all order. Then, in Sec.~\ref{sec:pdfs_beams} we will
explain the inclusion of parton distribution functions (PDFs) for hadron
beams as well as perturbatively calculable lepton PDFs for the
resummation of initial-state QED corrections in case of lepton
beams. Also, we discuss the simulation of beam effects in MC
generators, like beamstrahlung, beam energy spread, crossing angle and
polarization of beams. The next section,
Sec.~\ref{sec:partonshowers} is devoted to parton showers, the
simulation of the emission of up to a hundred strongly interacting
partons and a single-digit number of photons. We will touch upon weak
showers and discuss the combination of parton showers with fixed-order
simulations in so-called matching procedures. After that, in
Sec.~\ref{sec:hadronization} we explain how the final-state partons will
be combined into bound states of mesons and baryons via hadronization,
also called fragmentation. 

In particle physics, there is not a single MC generator that natively
accomodates all different aspects listed here, but there are a few
multi-purpose generators, a term which is not exactly well-defined,
sometimes restricted to those that simulate (parton shower and)
fragmentation, sometimes to multi-leg (parton-level) generators that
can generate arbitrary processes of (almost) arbitrary
multiplicity. The most well-known and most applied ones
are \textsc{Herwig}~\cite{Bahr:2008pv,Bellm:2015jjp,Bewick:2023tfi},
\textsc{Madgrapgh5\_aMC@NLO}~\cite{Alwall:2011uj,Alwall:2014hca,Frederix:2018nkq},
\textsc{Pythia}~\cite{Sjostrand:2006za,Sjostrand:2014zea,Bierlich:2022pfr},
\textsc{Sherpa}~\cite{Gleisberg:2003xi,Sherpa:2019gpd,Sherpa:2024mfk},
\textsc{Whizard}~\cite{Kilian:2007gr,Bredt:2022dmm,Reuter:2024dvz},
with a medium number of additional
generators like
e.g.~\textsc{Powheg-Box}~\cite{Nason:2004rx,Frixione:2007vw,Alioli:2010xd},
\textsc{CompHEP/CalcHEP}~\cite{Pukhov:1999gg,CompHEP:2004qpa,Belyaev:2012qa},
\textsc{Geneva}~\cite{Alioli:2012fc},
\textsc{Photos}~\cite{Barberio:1990ms,Barberio:1993qi,Davidson:2010ew},
\textsc{KKMC}~\cite{Jadach:2022mbe}
and many more that have some specific focus on the type of
interactions, the type of collider, or a particular type of
perturbative corrections. More details will be found in the sections
below. As described above, these generators cover all of particle
physics, so such a pedagogical review clearly has to focus on certain
key topics and cover some special aspects only in a very short way.

There are many excellent overview articles on MC event generators,
both from lectures~\cite{Skands:2012ts,Buckley:2011ms,Nason:2012pr},
and within the context of particle physics strategy
updates~\cite{Campbell:2022qmc}. Besides there extremely prominent
role in the data analysis and interpreation of LHC data, in the past
decade, MC generators have played a pivotal role as digital twins for
the planning of the next generation of ($e^+e^-$)
colliders~\cite{Aicheler:2012bya,Linssen:2012hp,ILC:2013jhg,Behnke:2013lya,FCC:2018evy,CEPCStudyGroup:2023quu,Berggren:2021sju,Altmann:2025feg,FCC:2025lpp,LinearColliderVision:2025hlt}
as well as muon
colliders~\cite{Accettura:2023ked,InternationalMuonCollider:2024jyv,InternationalMuonCollider:2025sys}.

%%%%%%%%%%%%%%%%%%%%%%%%%%%%%%%%%%%%%%%%%%%%%%%%%%%%%%%%%%%%%%%%%%%%%%%%%%%%%%%%
\section{The rationale of Monte Carlo generators}
\label{sec:mcgen}

\subsection{The need of Monte Carlo integration methods}

The main task of theoretical particle collider physics is to model
$2\to n$ scattering processes according to the differential cross
section formula:
\begin{equation}
  \label{eq:diffxsec}
  d\sigma_{\alpha\to\beta} =
  \frac{\left|\mathcal{M}_{\beta\alpha}\right|^2}{4 \sqrt{(p_{a,1}\cdot
      p_{\alpha,2})^2 - m^2_{\alpha,1}m^2_{\alpha,2}}}
    \left(\prod_{i=1}^n \frac{d^3q_\beta}{(2\pi)^3 2q_{\beta,i}^0}
    \right) (2\pi)^4 \delta^4((p_{\alpha,1}+p_{\alpha,2} - q_{\beta,1}-
  \ldots - q_{\beta,n})\; ,
\end{equation}
consisting of three components: the (squared) matrix element
$\mathcal{M}_{\beta\alpha}$, the kinematic flux factor in the
denominator and the integral over the kinematic final particle phase
space variables. For an $n$-particle final state, this integral has
dimension $3n-4$ (and two more variables in case of convolutions over
\begin{table}  
  \begin{center}
  \begin{tabular}{|l|r|r|r|}\hline
     Final state $X$ & particles &
     dim(partonic) & dim(with PDFs) 
      \\\hline
      $\mu^+\mu^-$ & 2 & 2 & 4
      \\
      $jjj$  & 3 & 5 & 7
      \\
      $\ell\ell bb$ & 4 & 8 & 10
      \\
      $\ell\ell bbj$ & 5 & 11 & 13
      \\
      $\ell\nu\ell\nu bb$ & 6 & 14 & 16
      \\
      \ldots & \ldots & \ldots & \ldots
      \\    
      $\ell\nu\ell\nu bb jjjj$ & 10 & 26 & 28
      \\\hline
  \end{tabular}
  \end{center}
  \caption{Dimensionality of the phase-space integration of several
    benchmark processes in particle physics. The left of the two
    dimensionality columns assumes a partonic process, while the
    rightmost column takes two variables from the energy fractions of
    the beam PDFs into account.}
  \label{tab:dimensionality}
\end{table}
initial state parton distribution functions, PDFs),
cf. \tabref{tab:dimensionality}. While there are explicit formulae
for low-dimensional phase
spaces~\cite{Byckling:1971vca}, for four or more final state particles
only numerical integration methods can be applied. Numerical methods
like Newton or Gauss integrations do not work very well: (i) the
dimensionality can become very high, with 6-10 particles for
complicated processes, (ii) the squared matrix elements has singular
or nearly singular structures due to propagator denomminators close to
their mass shells, and (iii) experimental selection cuts make the
topology of the integration manifold highly complicated. Hence, Monte
Carlo sampling is the only viable choice; besides being the only
feasible path for performing these high-dimensional integrations, it
also provides a method to generate events from a probablity density
given directly by~\eqref{eq:diffxsec}. Each kinematic configuration
will be generated with a probability given by $d\sigma/dq$ which acts
as a weight of the event. These weighted events can then be unweighted
by applying a veto algorithm, keeping events with a probability given
by the ratio of their weight divided by a maximal weight over all
phase space. Finally, these unweighted events (which can be given
uniform weight $w = 1$ or $w = \sigma_{\text{tot}}$) then are generated
with probabilities mirroring the ones in collider physics scattering
experiments.

The Monte Carlo method is based on the central limit theorem of
calculus that a (multi-dimensional) integral can be approximated by
volume of the integration domain, $V$, times the mean value of the
integrand, $\langle f \rangle$, where the error of the approximation
is given by the product of the volume $V$ and the variance, $\sigma$,
divided by the square root of evaluation points $N$: $I = V \cdot \langle
f \rangle \pm V \tfrac{\sigma}{\sqrt{N}}$. Here, $\langle f \rangle =
\tfrac{1}{N} \sum_{i=1}^N f(x_i)$, $\sigma^2 = \langle f^2 \rangle -
\langle f \rangle^2$. There are two ways to improve
the precision of a MC predictions, either to increase the number of
random number points (MC ``calls'') or to reduce the variance to bring
the MC error down. The most common method for variance reduction in
MCs is importance sampling, i.e. to sample integrand functions $f$ much
denser in peaked regions than in more ``unimportant'' domains of
$f$. This is achieved by a change of integration variables by means of
an invertible mapping (which is needed to calculate the Jacobian and
to convert between original and new variables):
\begin{equation}
  I = \int f \, dV = \int \frac{f}{g} dV = \langle \frac{f}{g} \rangle
  \pm \frac{1}{\sqrt{N}} \sqrt{\langle\left(\frac{f}{g}\right)^2 -
    \langle \frac{f}{g} \rangle^2} \qquad .
\end{equation}
Now, instead of sampling a function $f$ over the variables $dV$ one is
sampling the function $f/g$ over variables $g dV$. This can be used to
``map out'' e.g. Breit-Wigner resonance peaks or radiation kinematics
in QCD or QED that are enhanced for collinear or soft splittings. This
precision is encoded in the accuracy of the final MC prediction for the
total or fiducial cross section; this directly turns into the number
of weighted event (phase space calls) that are needed to generate a
single unweighted event (i.e. the unweighting efficiency becomes
closer and closer to 100\% the closer the phase space mapping turn the
integrand into a constant function).

Besides importance sampling, there is also stratified sampling where
the integration domain (e.g. in each dimension) is partitioned and
this partitioning is optimized. This can autoated over many different
dimensions and is codified in the \textsc{Vegas}
algorithm~\cite{Lepage:1977sw}. \figref{fig:s_t_channel} shows an
example of a one-dimensional stratification in the left panel. In
general, immportance sampling outperforms stratified sampling by far,
as stratified sampling samples functions where not where they are
largest, but where they are changing most rapidly. More about these
methods will be discussed in the following subsection.

For a general overview on Monte Carlo techniques in particle physics,
cf.~\cite{James:1980yn}. Finally, of course, MC techniques heavily
rely on high-quality pseudo-random number generators (RNGs), that need
to cover the high-dimensionl unit hypercube in a way as uniformly and
densely as
possible~\cite{James:1988vf,l1988efficient,Marsaglia:1990ig,Ferrenberg:1992zz,Luscher:1993dy,James:1993np,Vattulainen:1994uc,Shchur:1998wi}. Especially,
when phase space sampling is parallelized over many different cores,
cf. \subsecref{sec:parallelization}, it is crucial to make sure that
random number sequences on different parallel instances are not correlated. 

%%%%%

\subsection{Multi-channel sampling of high-dimensional phase spaces}

A generic phase space generator of an MC integrator consists of pseudo
random number generator (RNG) and a setup of mappings from the
variables of the unit hypercube to the kinematic moentum variables of
the final state particles. The simplest examples are completely
featureless and parameterize a cascade-like construction of the
collision energy from massless final-state momenta; this is encoded in
the \textsc{Rambo} algorithm~\cite{Kleiss:1985gy,Platzer:2013esa}.

Three avenues have been followed to more effectively sample
high-dimensional phase spaces, constrained by complicated experimental
cut selections: (i) importance sampling and 
subsection, (ii) stratified sampling have been mentioned already in
the last subsection. The third avenue is multi-channel sampling,
taking into account that the integrand function, the squared matrix
\begin{figure}
  \begin{center}
  \raisebox{-6mm}{\includegraphics[width=9cm]{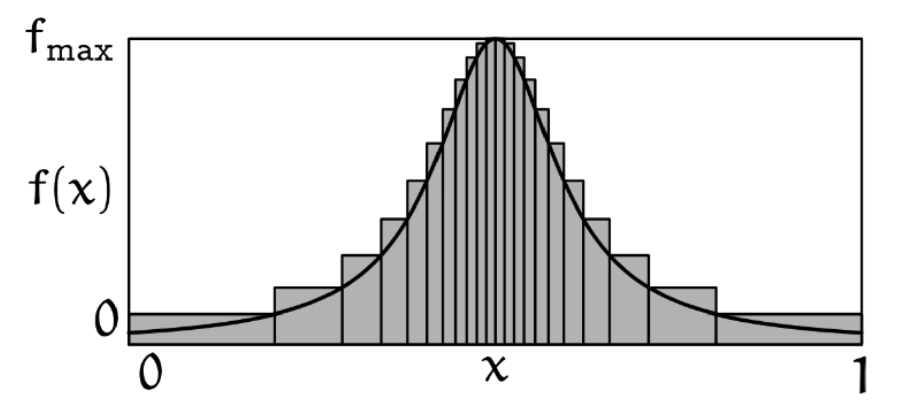}} \qquad
  \includegraphics[width=2.5cm]{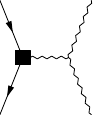} \qquad
  \includegraphics[width=2.5cm]{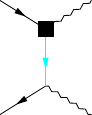}
  \end{center}
  \caption{Left: stratified sampling in one dimension; right:
    Multi-channel integration, two different kinematic channels for
    the process $e^+e^- \to W^+W^-$: $s$-channel diagram (left),
    $t$-channel diagram (right).}
  \label{fig:s_t_channel}
\end{figure}
element, has different features that cannot be mapped simultaneously
with the same variable transformation to a smoother function,
i.e. that the different resonance singularities are factorizable:
$g(x) = g(x_1) \cdot g(x_2) \ldots g(x_n)$. In such a case, the
integral can be split up into a sum over different ``phase space
channels''
\begin{equation}
  \label{eq:multi_channel}
  I = \int f(x) dV = \sum_i \alpha_i \int \frac{f(x)}{g(x)} g_i(x) dV
  \qquad .
\end{equation}
The channel weights, $\alpha_i$, constitute a partition of unity; in
an adaptive procedure, they can be optimize according to the
importance of different phase space channles within a single MC
integral~\cite{Kleiss:1994qy}. \figref{fig:s_t_channel} shows two
different phase space channels ($s$- and $t$-channel kinematics) for
the process $e^+e^- \to W^+W^-$ (right panel). Most of the art of
modern MC generators is in the clever choice of distribution into
phase space channels and their particular mappings. Examples of modern
phase space generators are \textsc{MadEvent}~\cite{Maltoni:2002qb},
the generator of \textsc{MadGraph5\_aMC@NLO}, where channels are
chosen according to heuristically dominant Feynmman diagrams. A
combination of importance sampling with physics-driven phase space
mappings, multi-channel adaptation and stratification is used in
\textsc{VAMP}~\cite{Ohl:1998jn}, the phase space sampler of
\textsc{Whizard}. There are many more examples. Several of these
algorithms can also handle overlapping singularities that cannot be
factorized. More recently, algorithms of neural importance sampling
based on invertible neural networks (INNs) have been implemented and
are becoming interesting competitors to more classical
algorithms~\cite{Heimel:2022wyj,Heimel:2023ngj,Bothmann:2020ywa}.

\subsection{Modern developments, validation and performance}
\label{sec:parallelization}

Besides very computing-intensive higher order matrix elements,
cf. \subsecref{sec:fixedorder}, phase space adaptation and sampling is
the major bottleneck of MC generators for multi-leg
processes. Parallelization over distributed or heterogeneous compute
architectures was a must to make complicated processes accessible to
the particle physics community. One aspect is the parallel evaluation
of matrix elements during phase space sampling, the other aspect is
the parallelization of the phase space sampling algorithm itself,
while event generation can be usually trivially parallelized by
dissecting the event samples into batches. For the parallel evaluation
of matrix elements, several options exist, e.g. using OMP
parallelization over different helicities or color flows in different
threads. A major step is to parallelize the phase space adapation
parallel in different phase space channels, using either Message
Processing Protocols like MPI~\cite{Brass:2018xbv} or graphics cards like
GPUs~\cite{Carrazza:2020rdn}.

MC generators need to be validated and during their development being
continuously tested for consistency, validity and efficiency. Modern
generators profit tremendously from cutting-edge programming paradigms
like encapsulation, object-oriented design, test-driven programming,
and validation paradigms like continuous integration and continuous
deployment. In the past, e.g. for LEP and Tevatron, different
generators have been validated by hand against other, while nowadays
automated software frameworks exist for continuous validation of
generator versions~\cite{Price:2025fzg}.

%%%%%%%%%%%%%%%%%%%%%%%%%%%%%%%%%%%%%%%%%%%%%%%%%%%%%%%%%%%%%%%%%%%%%%%%%%%%%%%%
\section{The hard scattering process}
\label{sec:hardscat}

In this section, we discuss the generation of matrix elements at tree
level and higher orders, \subsecref{sec:fixedorder}, which exhibit
infrared or mass singularities for (quasi) massless particles in
virtual or real corrections. In MC generators they have to be treated
with so called subtraction methods, which will be detailed in
\subsecref{sec:subtraction}. In \subsecref{sec:resum} we briefly comment
on combining MC simulations with analytic resummation to all orders.

\subsection{Fixed-order perturbation theory matrix elements}
\label{sec:fixedorder}

Fixed-order matrix elements are the main building blocks of the hard
scattering process: they are functions of the external momenta of the
incoming and outgoing particles as well as their quantum numbers like
helicities and color degrees of freedom. 

These amplitudes are either based on a diagrammtic expansion with a
\begin{figure}
  \begin{center}
    \includegraphics[width=.9\textwidth]{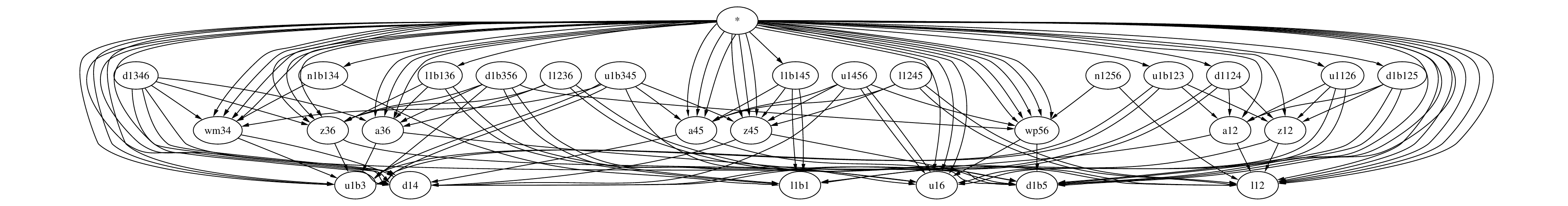}
  \end{center}
  \caption{Recursive algorithms for matrix elements reuse all possible
    subamplitudes and avoid all redundancies,
    from~\cite{Moretti:2001zz}.}
  \label{fig:recursion}
\end{figure}
subsequent common subexpressions elimination (CSE), or by using
recursive algorithms are avoiding redundancies from the very
beginning~\cite{Parke:1985ax,Berends:1987me,Berends:1987cv}. The
recursion, shown in \figref{fig:recursion} can be based on functional
programing paradigms~\cite{Ohl:2023bvv}. QCD quantum numbers need to
be accounted for, for which several algorithms exist, e.g. the color-flow
formalism~\cite{Maltoni:2002mq,Kilian:2012pz,Sjodahl:2012nk,Sjodahl:2014opa,Ohl:2024fpq},
as well as relative sign of subamplitudes according to Fermi
statistics. This can be fully automated recursively, but is quite
intricate for fermion-number non-conserving interactions
(e.g. supersymmetric models and models with Majorana
fermions)~\cite{Ohl:2002jp}.

Automated generators of fixed-order matrix elements:
many of these matrix element generators are historically attached to a
MC generator framework (listed in parentheses below), but most if not
all of them are in principle standalone and could be used
interchangeably within different MC frameworks:
\textsc{Alpha}~\cite{Caravaglios:1995cd} (\textsc{Alpgen}), 
\textsc{Amegic/Comix}~\cite{Krauss:2001iv,Gleisberg:2008fv} (\textsc{Sherpa})
\textsc{MadGraph}~\cite{Stelzer:1994ta} (\textsc{MadGraph5\_aMC@NLO}
\textsc{O'Mega}~\cite{Moretti:2001zz} (\textsc{Whizard})
\textsc{CompHEP/CalcHEP}~\cite{Pukhov:1999gg,CompHEP:2004qpa,Belyaev:2012qa}
(same MC name), and many more.

Basically all automated matrix element generators provide code in a
compilable computer language like \texttt{C++} or
\texttt{Fortran}, or more recently with \texttt{Cuda} support for
GPUs~\cite{Hagiwara:2013oka,Carrazza:2021gpx,Carrazza:2020qwu,Bothmann:2021nch},
which is then compiled and linked into dynamic libraries. For
complicated processes, this process code and the binaries can become
quite large. Alternatives are to transfer matrix elements into
bytecode instructions (wavefunctions to be multiplied with vertex
factors and propagators) which are interpreted by a single binary,
where such an interpreter acts as a ``virtual
machine''~\cite{ChokoufeNejad:2014skp}. 

For the Standard Model (SM), there is a well-defined subset of building
blocks in terms of Lorentz representations of external particles and
propagators, Lorentz structures for vertices, and all that. Extending
the SM towards higher-dimensional operators in a framework like SM
effective field theory (SMEFT) or studying an arbitrary framework
beyond the SM (BSM) necessitates an almost infinite plethora of
building blocks. In the past, BSM models needed to be hard-coded and
the validation of these implementations was tedious (a famous example
is the minimal supersymmetric SM, MSSM~\cite{Hagiwara:2005wg}). The
situation was vastly improved with Lagrangian-level tools, where BSM
models can be typed in using a textbook-like format and get
automatically exported to MC generators. Examples are
\textsc{LanHEP}~\cite{Semenov:2008jy},
\textsc{SARAH}~\cite{Staub:2013tta} or
\textsc{FeynRules}~\cite{Alloul:2013bka}. Now, still for each of these
tools the interface to each specific MC generator needs to be
validated and maintained, cf. e.g.~\cite{Christensen:2010wz}. Hence,
an intermediate layer has been invented, the Universal Feynman Output
format~\cite{Degrande:2011ua,Darme:2023jdn}, such that now each
Lagrangian tool and each MC generator only has to validate and
maintain a single (UFO) interface. Connected to this interfaces there
are convenience tools for automated parameter scans, automated
calculations of decay widths and branching ratios.

%%%%%

\subsection{Higher orders and subtraction methods}
\label{sec:subtraction}

Higher order corrections in a perturbative series in quantum field
theory contain either real or virtual corrections. At next-to-leading
order (NLO), real corrections consist of the square of the
real-emission diagrams and the interference of the tree-level with the
one-loop amplitude. At the next order, NNLO, there are double-virtual
contributions, the interference of the tree-level with the two-loop
amplitude, double-real corrections (the square of double-emission
diagrams) and real-virtual corrections (intereferences of loop
diagrams with diagrams with a single real emission). Real-emission
diagrams are simply tree-level diagrams of higher mulitplicity, while
loop diagrams can be constructed by their analytic properties from
tree-level diagrams by so-called unitarity cut
methods~\cite{Ossola:2006us,Mastrolia:2012bu}. The loop integrals over
internal momenta can then be algebraically reduced to basic ``master
integrals''~\cite{Passarino:1978jh,Davydychev:1991va,Denner:2005nn}.
Due to the relative size of the QCD coupling and the QED/weak coupling
$0.1 \alpha_s \sim \alpha$, for hadron colliders like the LHC
generically NLO EW corrections are considered to be of the same
importance as NNLO QCD corrections, explaining the focus on the QCD
corrections. For lepton colliders, this is almost reversed as most
signal processes are free from QCD corrections and NLO or NNLO QED/EW
corrections dominate the theory uncertainties. 

An important concept of quantum field theories with massless particles
is that of infrared safety: only observables that are defined in a way
such that they are not affected by soft or collinear emissions of
massless particles, yield finite results in perturbation theory. This
is encoded in the Kinoshita-Lee-Nauenberg (KLN)
theorem~\cite{Kinoshita:1962ur,Lee:1964is}. This is closely related to
the concept of QCD jets, which are bundles of strongly interacting
particles defined by a geometric procedure in phase space in order to
ensure infrared safety, a so called jet algorithm, e.g. the
Cambridge-Aachen, $k_T$ or anti-$k_T$
algorithm~\cite{Sterman:1977wj,Brown:1991hx,Catani:1993hr,Dokshitzer:1997in,Cacciari:2008gn,Cacciari:2008gp,
Cacciari:2011ma}. For more  details cf. the section on ``Perturbative 
QCD''~\cite{Heinrich:2025ylf}. The emission of soft or/and collinear
QCD particles in both virtual and real corrections generate mass
singularities that cancel each other in infrared quantities. In
analytic calculations, these singularities are regularized in an
algebraic manner, e.g. dimensional regularization. This is not
possible for MC generators. Until ca. 2000, complicated partitions of
phase space (``phase space slicing'') had been applied, where for each
process at higher order the independence of the final results on the
partitioning of phase space (slicing parameters) had to be proven. By
the end of the 1990s, a more sophisticated framework had been
developed in terms of subtraction algorithms:
\begin{equation}
  \label{eq:nlo_subtraction}
  d\sigma_{2\to n}^{\text{NLO}} = d\Phi_n \left[ \mathcal{B}_n +
    \mathcal{V}_n + \mathcal{B}_n \otimes S\right] + d\Phi_{n+1}
  \left[ \mathcal{R}_{n+1} - \mathcal{B}_n \otimes dS \right]
\end{equation}
from the real emissions $\mathcal{R}_{n+1}$ to a $2\to n$ Born process
$\mathcal{B}_n$, their soft/collinear singular parts $dS$ are subtracted in
a way that they can be analytically (or semi-numerically) be
integrated over the phase space variables of the QCD radiation. This
enables an analytic integration of the singular parts which can be
added to the virtual components $\mathcal{V}_n$ of a higher-order
calculation, such that each component is separately infrared
finite. Subtraction is based on the property that
in the soft and/or collinear limit, quantum field theoretic $n+1$
particle amplitudes factorize into $n$ particle amplitudes times a
$1\to 2$ splitting function. In addition, also the $n+1$ particle
phase space factorizes into the $n$ particle phase space times a
one-particle radiation phase space. The two main subtraction
algorithms are Catani-Seymour (CS)~\cite{Catani:1996vz,Catani:2002hc}
and Frixione-Kunszt-Signer (FKS)
subtraction~\cite{Frixione:1995ms,Frixione:1997np}, while there
are also more numerical approaches~\cite{Nagy:2003qn}. 

A typical MC generator that provides automated NLO
calculations/simulations then consists of a framework of automatically
generating all the subtraction terms with the corresponding
phase-space setup, providing tree-level matrix elements for the Born
process, the real emissions, the color- and spin-correlated matrix
elements for the subtraction terms in the soft and collinear regions,
respectively. Either the MC generator has its own one-loop matrix
element generator or receives the virtual amplitudes from a one-loop
provider (OLP). These tools consist of libraries that contain the
tensor reduction into master integrals and libraries for the scalar
master
integrals~\cite{Ossola:2007ax,Peraro:2014cba,Denner:2014gla,Denner:2016kdg,Hirschi:2016mdz}. The
interface between MC generators and OLP tools 
has been standardized in the Binoth Les Houches Accord
(BLHA)~\cite{Binoth:2010xt,Alioli:2013nda}. Examples of these
interfaces can be found in~\cite{Biedermann:2017yoi,Bredt:2022dmm}.

As discussed above, unitarity cuts, tensor reduction to master
integrals and libraries of all needed master integrals have triggered
the famous ``NLO revolution'', the automation of NLO matrix element
generation (and their usage via subtraction formalism in MC
generators) starting from ca. 2010. Automated one-loop generators are
\textsc{GoSam}~\cite{GoSam:2014iqq,Braun:2025afl}, \textsc{MadLoop}~\cite{Hirschi:2011pa},
\textsc{OpenLoops}~\cite{Cascioli:2011va,Buccioni:2019sur} and 
\textsc{Recola}~\cite{Actis:2016mpe}.
Using these OLP tools, NLO QCD+EW, i.e. full NLO SM processes have
been automated in
\textsc{MadGraph5\_aMC@NLO}~\cite{Alwall:2014hca,Frederix:2018nkq},
\textsc{Sherpa}~\cite{Biedermann:2017yoi,Sherpa:2024mfk} and
\textsc{Whizard}~\cite{Rothe:2021sml,Bredt:2022nkq}. 

This NLO paradigm can be in principle extended towards (automated) NNLO
tools, however, two-loop integrals, especially with many legs and/or
many different internal mass scales are highly complicated and not
automated yet, and though there is tremendous progress for subtraction
schemes at NNLO for QCD, there is no completely automated NNLO
subtraction scheme for QCD and EW interactions. The work on general
subtraction schemes for NNLO QCD is a very active field of
research~\cite{Gehrmann-DeRidder:2005btv,Catani:2007vq,Czakon:2010td,Czakon:2011ve,Boughezal:2011jf,Currie:2013vh,Cacciari:2015jma,Boughezal:2015eha,Boughezal:2015dva,Gaunt:2015pea,DelDuca:2015zqa,Caola:2017dug,Caola:2018pxp,Magnea:2018hab,Magnea:2018ebr,Delto:2019asp,Magnea:2020trj,Bertolotti:2022aih,Bertolotti:2022ohq,Braun-White:2023sgd,Braun-White:2023zwd,Devoto:2023rpv,Fox:2023bma,DelDuca:2024ovc,Devoto:2025kin}. More details
can be found in~\cite{Heinrich:2025ylf}. In addition, mixed QCD-EW
corrections have become possible~\cite{Heller:2020owb,Bonciani:2021zzf,Buccioni:2022kgy,Armadillo:2022bgm}.
As of now, there is no automated matrix element generator for NNLO
QCD, but there are libraries of the most important processes like
Higgs production, Drell-Yan, dibosons and top pairs,
cf. e.g.~\cite{Hartanto:2019uvl,Badger:2021ega,Badger:2021nhg,Badger:2022ncb,Badger:2024sqv,Chawdhry:2020for,Chawdhry:2021mkw}
or event shapes in $e^+e^-$~\cite{DelDuca:2016ily}. The
\textsc{Matrix} framework~\cite{Grazzini:2017mhc} combines a larger
list of NNLO QCD processes. NNLO
EW corrections only exist for very few processes, e.g. dominant
corrections for $e^+e^-\to WW$~\cite{Actis:2008rb} and the full
corrections $e^+e^- \to ZH,WW$~\cite{Freitas:2022hyp} and attempts for
the automation are still in its infancy, note
however~\cite{Chen:2022dow}. For simulation tools for low-energy
experiments like MUonE, the situation becomes simpler as all leptons
can be treated massive and only soft singularities remain: this makes
the subtraction much simpler. This has been used in the
\textsc{McMule} tool which aims at NNLO
accuracy~\cite{Banerjee:2020rww}. An extensive overview for tools for
low-energy experiments can be found here~\cite{Aliberti:2024fpq}.

There are several places in an NLO or NNLO calculation which could
produce negative weights: (i) NLO or higher order fits of PDFs need
not necessarily be positive definite any more, (ii) outside the strict
soft or collinear limits, cancellation between the real matrix
elements and their soft-collinear approximation can turn negative, and
(iii) there regions of phase space where the cancellation of
subtractions and their integrated terms is numerically
imperfect. Negative weights normally average each other out with
positive weights in binned distributions, but they hamper the
efficiency of NLO event generation, as for several negative weights
events a corresponding number of positive weight events need to
generated. This greatly enlarges the number of needed events to reach
a current precision. Therefore, algorithms to reduce the number of
negative weights, e.g. by more cleverly grouping them or resampling
techniques~\cite{Frederix:2020trv,Nachman:2020fff,Andersen:2020sjs,Danziger:2021xvr,Andersen:2021mvw,Frederix:2023hom,Andersen:2023cku}.

A special case is the top quark when produced close to threshold, as
it exhibits effects of a quasi-bound state similar to hadronization
effects (cf. \secref{sec:hadronization}), which, however, need to be
handled in the framework of the hard fixed-order process. In MC
generators, this has been first addressed for the top threshold in
$e^+e^-$ collisions~\cite{ChokoufeNejad:2016qux,Bach:2017ggt}, while
recently it has been realized that this effect is experimentally
visible also at LHC~\cite{Fuks:2025wtq}.

\subsection{Combining fixed order with resummation tools}
\label{sec:resum}

For many aspects, fixed-order calculations within a perturbative
series are not sufficient to achieve a precision goal for
theoretical/MC predictions. Due to large kinematic scale separations
(e.g. typical jet energy scales of hundreds of GeV at the LHC and the
hadronization scale of $\Lambda_{\text{QCD}} \sim$ 1 GeV, large
logarithms occur, such that for a small coupling $\alpha_s \ll 1$
the product $\alpha_s \log Q_1/Q_0 \sim 1$ or even larger than one. A
typical example are QCD Sudakov logarithms which are e.g. addressed in
\secref{sec:partonshowers}. Especially for more inclusive quantities
like total or single-differential cross sections, it is possible to
resum such logarithms to all orders, where such resummation is
performed using methods from effective field theories (EFTs) which
have the same low-energy behavior than the full theory (here e.g. QCD)
but are much simpler in the ultraviolet. Such theories are
e.g. soft-collinear EFT (SCET). While such resummation calculations
very often have to be done in a process-specific manner, there are
several tools that have been developed that take care of certain
universal features of such resummations.
Examples of interfacing resummation tools to fixed order simulations,
are~\cite{Banfi:2004yd,Baberuxki:2019ifp}, very often in the context
of jet or event shape variables. For more information see the section
on jets in $e^+e^-$~\cite{Stagnitto:2025air}. 

Another example of such large logarithms are electroweak (EW) Sudakov
logarithms~\cite{Denner:2000jv,Denner:2001gw}: these originate from
incomplete cancellations between virtual and real corrections as
initial or final states are not EW singlet states or because the event
selection do not include explicit EW real radiation. For TeV-scale
processes like the production of EW bosons at the LHC or basically any
process at a future 3-10 TeV muon collider, these logarithms are quite
large and can reduce the Born cross section by 20 to 100 per
cent. Given the external lines of a process, tools as add-ons to MC
generators can automatically generate such EW Sudakov logarithms and
dress MC processes with
them~\cite{Bothmann:2020sxm,Pagani:2021vyk,Lindert:2023fcu,Denner:2024yut}.
Applications of EW Sudakov resummations for LHC can be found
in~\cite{Bothmann:2021led,Pagani:2023wgc} and in~\cite{Ma:2024ayr} for
the muon collider.

%%%%%%%%%%%%%%%%%%%%%%%%%%%%%%%%%%%%%%%%%%%%%%%%%%%%%%%%%%%%%%%%%%%%%%%%%%%%%%%%

%%%%%%%%%%%%%%%%%%%%%%%%%%%%%%%%%%%%%%%%%%%%%%%%%%%%%%%%%%%%%%%%%%%%%%%%%%%%%%%%
\section{Parton distribution functions and beam simulations}
\label{sec:pdfs_beams}

Most beam particles within collisions in particle physics are not
pointlike particles: this is definitely true for proton collisions,
where the beams consists of bound states of quarks and gluons, but
also of leptons like electron and muons. The latter undergo QED or weak
interactions and appear at high energies as well as a collinear bunch
of electromagnetically or weakly interacting particles. In this
section, we will discuss the inclusion of parton distribution
functions in MC generators, first for proton beams in
\subsecref{sec:proton_pdf}, then for lepton beams in
\subsecref{sec:ele_pdf}, and finally comment on aspects like beam
spectra of lepton or photon beams, crossing angles and polarization in
\subsecref{sec:beamsim}.

\subsection{Parton distribtion functions of the proton}
\label{sec:proton_pdf}

Parton distribution functions (PDFs) of quarks and gluons within
protons, modelled in a Lorentz frame fully collinear with the incoming
beams, are being fitted from data, from deep-inelastic scattering 
experiment like the electron-proton collider HERA or from
proton-proton collisions like LHC. The errors from these fits,
especially for the gluon PDF which is the most important for many LHC
processes, are for many measurements one of the largest sources of
uncertainties. There are several different collaborations each doing
independent fits, like ABM~\cite{Alekhin:2013nda},
CTEQ~\cite{Nadolsky:2008zw,Pumplin:2002vw,Gao:2013xoa}, HeraPDF~\cite{H1:2015ubc},
MSTW/MMHT~\cite{Martin:2009bu,Harland-Lang:2014zoa}, 
NNPDF~\cite{NNPDF:2014otw,NNPDF:2017mvq}, cf. as
well~\cite{Butterworth:2015oua,Cruz-Martinez:2024cbz}. For the MC
generator, the details of these different fits do not really matter as
long as these PDF fits are available in a fast and standardized
manner: this is done in a library like LHAPDF~\cite{Buckley:2014ana}
(standardized through an effort during Les Houches workshops and the
NLO-MC BLHA interface or LHA/LHE event formats). The generator calls
the PDF as function of the parton flavor, the energy fraction of the
incoming beam particle (``Bjorken $x$''), the factorization scale, the
perturbative QCD order at which the PDF had been performed and certain
validaty boundaries of the fits. Generally the parton flavor has 11
components, the gluon and five quark and anti-quark components each,
with the top (anti-)quark content assumed to be zero. Many PDFs now
include the photon content inside the proton~\cite{Manohar:2016nzj},
which becomes important for high-charge ion collisions generating a
huge photon flux. More recently, also lepton content in proton PDFs
have been considered, e.g. from photon-to-lepton pair splitting at
NNLO QED level~\cite{Buonocore:2020erb}. 

These PDF fits are provided as grids of certain values with specific
interpolation and extrapolation routines between and beyond. Though
this looks trivial from the point of the MC generator, it is crucial
that these calls to PDF values are fast and efficient: the PDFs are
not only used for convolutions with matrix elements, but also as
ratios of PDFs in the construction of subtraction terms and in Sudakov
factors of initial-state parton showers.

%%%%%

\subsection{Underlying event and multi-parton interactions}
\label{sec:underlying}

Very briefly, multi-parton interactions (MPI, do not confuse with the
Message Processing Interface) in \subsecref{sec:parallelization}) is
specific to hadron collisions: there can be more than one parton
undergoing a hard interaction. For the MC generator this is a double
invocation of its routines; the main new feature is to access the
probabilities for the double-parton splitting out of a hadron, for
which special PDFs exist that need to be interfaced to the
generator. Secondly, there is the ``underlying event'' which collects
all additional effects that are connected to the initial state: the
beam remnant which is either a color-(anti)sextet in the case of a
(anti-)quark or a color octet in the case of a gluon as a parton
entering the hard scattering process. The beam remnant also undergoes
QCD radiation simulated by the parton shower and needs to be
transferred into a system that can be processed by the
hadronization, including the initial conditions for the kinematics of
the beam remnant. In addition, also the generation of ``intrinsic''
$p_T$ is counted with the underlying event: while PDFs are fitted in
the strictly collinear limits, the parton splitting can also be
generate transverse momentum (as it does e.g. in transverse-momentum
dependent PDFs, TMDs). This ``intrinsic'' transverse momentum
contributes to the $p_T$ distribution of jets and electroweak
particles at hadron colliders in the lowest bins of a few to 10-15
GeV. For more details see the discussions in Sec. 10
of~\cite{Sjostrand:2006za}.

%%%%%

\subsection{Electron/lepton parton distribution functions}
\label{sec:ele_pdf}

At high enough lepton beam energies, $Q_\ell$, even for for the
smaller QED coupling constant, $\alpha \sim \alpha_s / 10$, the
quantity $\alpha \log{Q_\ell^2/m_\ell^2}$ becomes of an order where
resummation is necessary. In contrast to QCD partons, the initial
conditions need not be fitted from experimental data, but can be
perturbatively calculated from first principles. The result is an
object that gives the probability of finding a ``partonic'' lepton,
anti-lepton or photon within the beam photon with a certain energy
fraction $0 \lesssim x \lesssim 1$, $f_\ell(x,Q_\ell)$. These objects
historically have been called lepton structure functions, but nowadays
are mostly called lepton PDFs: structure functions are considered to
be based on kinematic approximations, while lepton PDFs are
field-theoretic objects that obey renormalization group or DGLAP
equations. 

In the derivation of lepton PDFs, soft photons can be resummed into a
compact formula to all orders~\cite{Gribov:1972ri,Gribov:1972rt}, most
easily using the formalism of Mellin transforms, to
which hard-collinear corrections have been added to first,
second~\cite{Kuraev:1985hb,Nicrosini:1986sm} and third 
order in
$\alpha$~\cite{Skrzypek:1990qs,Cacciari:1992pz,Arbuzov:1999cq}. All of
these formulae resum the leading-logarithmic (LL) corrections. In the past
years, anticipating the needs of a future $e^+e^-$ Higgs factory, a
lot of effort has been put into resumming the NLL
terms~\cite{Frixione:2019lga,Bertone:2019hks,Bertone:2022ktl,Frixione:2012wtz},
and calculating the fixed-order NNLO QED
contributions~\cite{Schnubel:2025ejl} to these PDFs. For a review on
these topics, cf.~\cite{Frixione:2022ofv}. The biggest phenomenogical
differences to QCD PDFs are (1) that the rise in the infrared is less
steep for lepton PDFs as QED is not asymptotically free so the rise
simply comes from the masslessness of the degrees of freedom, and (2)
that the lepton PDFs exhibit an integrable singularity in the limit of
Bjorken-$x$ $x \to 1$ (while QCD PDFs vanish in this limit). For the
MC generator this singularity results in a huge challenge regarding
numerical stability, which is manageable in convolutions with
tree-level matrix elements, but becomes quite intricate when combined
with subtraction terms for higher order calculations. To provide
a stable automated framework for NLO EW calculations in MC generators
is an active field of research.

Note that there are several dedicated programs for NLO EW corrections
for specific processes, that include the effects of these lepton PDFs
at LL or NLL accuracy~\cite{Denner:2002cg,Beneke:2007zg}. In addition,
it is, of course, also possible to do NLO EW calculations in MC
generators for lepton colliders using massive initial state leptons,
e.g.~\cite{Bredt:2022dmm,Bredt:2025zxc}.

At very high energies, e.g. future 10 TeV parton level colliders like
the muon collider MuC or FCC-hh, one can consider the full SM as
partons, generalizing the concept of QED lepton PDFs to EW lepton
PDFs. As EW interactions are chiral, partons need to be considered
polarized, and all of them are coupled to the DGLAP equations of the
full SM. Counting all degrees of freedom yields 59 in the SM
(including interferences induced by EW symmetry
breaking)~\cite{Han:2020uid,Garosi:2023bvq}. These EW PDFs have been
put into the same framework like proton PDFs by numerically solving
and interpolating those coupled DGLAP equations; they are the initial
state counterparts of the full soft-collinear realm of the SM to the
EW parton showers in \subsecref{sec:qedshower}. In this framework they
are available in the MC generators (cf. e.g. to the framework
in~\cite{Dahlen:2025udl}).

%%%%%

\subsection{Beam simulations of lepton and photon colliders}
\label{sec:beamsim}

There are several aspects of the physics at lepton colliders that are
different from hadron colliders. Electron-positron colliders at high
energies and/or high luminosities exhibit beamstrahlung, classical
radiation caused by the electromagnetic fields from the other bunch
shortly before the collisions. This has to be taken into account in
order to carefully plan radiation occupancies in detectors under
development and to estimate systematic uncertainties for
reconstruction and analysis. In addition, these effects lead to
luminosity smearing which is a convolution of three different effects:
the natural beam energy spread resulting from the machine design (of
the order of 0.02 - 0.1 \%), beamstrahlung determined from  classical
electrodynamics which depends on the beam optics in front of the
collision point, especially the final focus magnets, and QED initial
state radiation. There exist dedicated simulation tools developed in
accelerator physics like \textsc{Guinea
  Pig}~\cite{Schulte:1999tx,Rimbault:2007wfy},
\textsc{Cain}~\cite{Chen:1994jt},  
\textsc{Fluka}~\cite{Ferrari:2005zk} or
\textsc{XSuite}~\cite{Iadarola:2023fuk}.  
At synchrotrons like LEP, CEPC and
FCC-ee this leads mostly to a Gaussian beam spread, while for linear
colliders like LCF, ILC, CLIC or C3 it leads to sizeable effects that
need to be taken into account. While synchrotrons mostly exhibit
transverse polarization of the electrons due to the Sokolov-Ternov
effect from synchrotron radiation, linear colliders allow for
longitudinally polarized beams~\cite{Moortgat-Pick:2005jsx}. All major
multi-leg MC generators allow for the simulation of longitudinally
polarized beams (also with polarization fractions in event generation
different from 100\%), while the \textsc{Whizard} framework allows
for arbitrary polarization (i.e. mixtures of longitudinal and
transverse polarizations or arbitrary spin quantization axes, in
general a completely general initial state spin density matrix).  

For beam simulations and beam spectra there are three levels of
sophistications: (i) the simulation of a two-sided or one-sided
Gaussian beam spread, (ii) a parameterized spectra and (iii) an MC
generator based on a two-dimensional histogrammed fit. Option (i) is
available in almost all MC generators for $e^+e^-$ colliders and is
very likely sufficent for muon colliders and synchrotrons at lower
energies (while FCC-ee at 365 GeV shows deviations from the Gaussian
profile). The parameterized spectra (option ii) assumes that the beam
spectra of the $e^-$ and $e^+$ can be factorized and each of them can
be described by a smeared delta peak and a polynomial tail, $D_i(x)
\approx \alpha \delta_\epsilon (x) + \gamma_i x^{\alpha_i}
(1-x)^{\beta_i}, i =1,2$. This formalism is sufficient for low-energy
ILC-type and C3-type machines (with energies $\lesssim 500$ GeV), and
it is implemented in~\cite{Frixione:2021zdp} and the first version
in~\cite{Ohl:1996fi}. Option (iii) is the most versatile approach: it
does not assume that the effects from the two beams can be
factorized. For this, it uses a two-dimensional grid fitted to the
low-statistics outputs of the machine simulation tools listed
above. To compensate artifacts of this low statistics a smoothing with
a Gaussian filter is applied. In addition, the interior of the square
$(x1,x2) \sim ([0,1]\times[0,1])$ and the boundary have to be fitted
independently in order to avoid artificial beam energy spreads. This
formalism is encoded in the \textsc{Circe2}
algorithm~\cite{Ohl:1996fi} included in the
\textsc{Whizard}~\cite{Kilian:2007gr}. Such a description is needed
for drive-beam accelerators like CLIC, plasma-wakefield accelerators
and also photon colliders based on Compton backscattering of electrons
from laser photons. Of course, it covers also ILC machine setups, and
it has been applied to the beam simulation of CEPC and FCC-ee (where
all four energy stages and all four interaction points).

%%%%%%%%%%%%%%%%%%%%%%%%%%%%%%%%%%%%%%%%%%%%%%%%%%%%%%%%%%%%%%%%%%%%%%%%%%%%%%%%

\section{Parton Showers}
\label{sec:partonshowers}

In this section we discuss the generation of QCD, electromagnetic and
weak radiation via parton showers and the matching to fixed-order
matrix elements. Additional information can be found on the section on
jets at electron-positron colliders~\cite{Stagnitto:2025air}. 

\subsection{Parton showers: rationale and implementation}
\label{sec:shower}

Typical events at high-energy hadron collider contain up to a hundred
different final state particles (e.g. constituents of jets), and even
high energy lepton collider events contain 10-20 particles. There are
two main obstacles to simulate such multiplicites with the methods of
the last two sections: (i) the dimensionality of phase space extends
the capacity of MC methods, and (ii) matrix element generation even
for tree-level processes hits its limitations for 15-20 external
particles. On the other hand, it is also not really necessary to
simulate most of these components with complete matrix elements, as
these emissions are dominated by soft and/or collinear emissions. We
can give here only a short introduction into parton showers, for a
general overview cf. the lecture
notes~\cite{Hoche:2014rga}. Historically, the concept of parton
showers have been developed together with the first physics results on
jets at the PETRA storage ring at DESY. As for subtraction terms in
fixed order calculations (cf. \subsecref{sec:subtraction}), parton
showers are based on the  structure of quantum field theories in the
infrared where $n+1$ particle amplitudes factorize into the
convolution of $n$ particle amplitude times splitting function. In the
soft-/collinear limit, not only amplitudes factorize, but also the
phase space: $d\Phi_{n+1} \approx d\Phi_n \cdot d(\Phi_{\rm rad})$. In
this limit, emissions become independent of each other and can be
exponentiated, where the exponential gives the total probability for
any number of emissions to happen. The inverse of this exponential,
the Sudakov form factor, then yields the no-emission probability:
$\Delta(t,t_0) = \exp \left[ - \int_{t_0}^t \tfrac{dt'}{t'}
\int_{z_-}^{z_+} dz \tfrac{\alpha_s(z,t)}{2\pi} P(z,t) \right]$.
It can be used to implement a veto algorithm: a random
number is drawn and compared to the Sudakov form factor to decide
whether to do an emission or not. $t$ is the ``shower time'' variable:
different parton shower algorithms use different kinematic variables,
the emission angle $\theta$, the parton virtuality $Q$, transverse
momentum $p_\perp$. Final state showers perform a forward evolution in
shower time from large to low scales, while initial-state showers do a
backward evolution. For the initial-state shower, the Sudakov form
factor contains ratios of the PDFs, as one has to divide out the PDF
of the Born parton without emission and multiply with the PDF of the
parton before the emission. The variable $z$ is a kinemmatic variable
of the splitting, usually the energy fraction of the original parton,
while $P(z,t)$ is the DGLAP splitting function, which obeys a QCD
evolution equation. An important concept of parton showers is color coherence by (large
angle) soft emissions: this is connected to angular ordering, where
fist emissions have to be carried out at largest angles. This
guarantees the dominant color coherence effects. 
There exist many different parton shower algorithms, based on
different concepts of the shower evolution or splitting variable, the
included order, the recoil scheme etc. Some examples
are~\cite{Nagy:2020rmk,Knobbe:2023njd,Kilian:2011ka,Masouminia:2021kne}.
Multi-purpose event generators, that have their
own parton shower implementations, are
\textsc{Herwig}~\cite{Bahr:2008pv,Bewick:2023tfi}, \textsc{Pythia}~\cite{Bierlich:2022pfr},
\textsc{Sherpa}~\cite{Sherpa:2019gpd,Sherpa:2024mfk} and
\textsc{Whizard}. There are other, stand-alone shower implementations
like \textsc{Ariadne}~\cite{Lonnblad:1992tz},
\textsc{Deductor}~\cite{Nagy:2007ty} etc.

The shower ``evolution'' is a unitary evolution, i.e. the parton
shower emissions (when being fully inclusive, summing over all
emission multiplicities and integrating over all of phase space) do
not change the total cross section of the Born process (total cross
sections, however, have to change when matching parton shower
emissions to emissions from hard matrix elements,
cf. \subsecref{sec:matching}. Parton showers do resum Sudakov
logarithms~\cite{Sudakov:1954sw} (which had been discovered for
analogous QED emissions), which make them the ``exclusive part'' of
resummation algorithms, cf. \subsecref{sec:resum}. Due to their
probabilistic (Markovian) nature, the assessment of theoretic
uncertainties is much harder than for an analytic resummation. The
past decade has seen a lot of progress quantifying the uncertainties
of parton showers beyond just running different shower algorithms and
comparing the differences. Angular-ordered
showers~\cite{Marchesini:1983bm,Marchesini:1987cf,Gieseke:2003rz} play
an important role, and with a careful bookkeeping of the kinematics of the
emissions~\cite{Bewick:2019rbu}, such angular–ordered showers maintain
NLL accuracy accuracy for global observables, i.e. not too
differential observables. These efforts of the past years, many
efforts have been devoted to go beyond the quasi-classical
approximation of leading splitting kernels (i.e. independing emissions
at the LL order) to achieve next-(next-)to-leading logarithmic 
accuracy~\cite{Bewick:2019rbu,Forshaw:2020wrq,Herren:2022jej,Hoche:2017iem,Dulat:2018vuy,Dasgupta:2020fwr,FerrarioRavasio:2023kyg,vanBeekveld:2024wws,Preuss:2024vyu,FerrarioRavasio:2025soo}.
This is achieved by including more exact kinematics, by incorporating
different recoil schemes and to include higher orders in spin and color
correlations. It is technically quite challenging to prove NLL or NNLL
accuracy of parton showers, using known results from analytic
resummation. \figref{fig:shower_matching} shows in the left plot a
comparison of two different parton showers (\textsc{Alaric} and
\textsc{Dire} from the \textsc{Sherpa} framework) with LEP data.

\subsection{Matching fixed-order processes to parton showers}
\label{sec:matching}

With dedicated approaches, it is possible to preserve the accuracy of the hard
scattering process even at higher perturbative orders. At NLO, there exist
techniques to match NLO calculations with
parton showers and obtain NLO+PS predictions that preserve NLO accuracy for
integrated distributions and Born-like quantities. In essence, these algorithms
provide a way to exactly describe one hard emission and fill the remaining phase
space with a shower, avoiding possible double counting.
\begin{figure}
  \begin{center}
    \includegraphics[width=.4\textwidth]{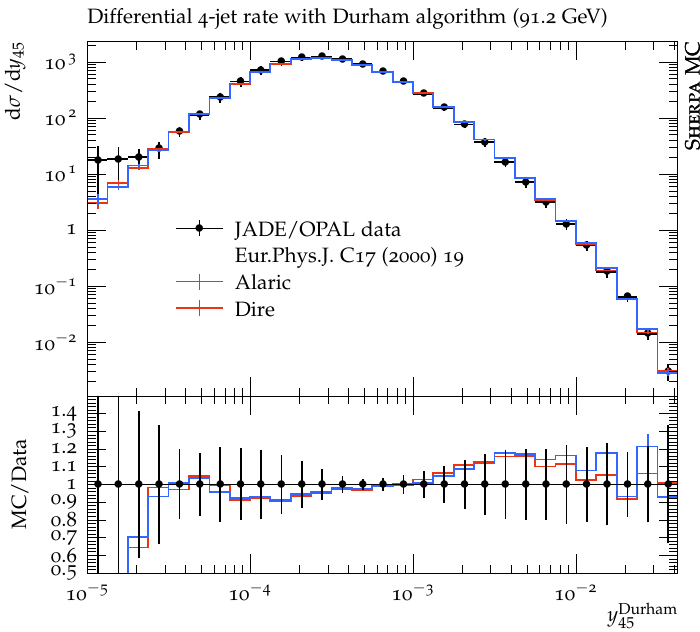}
    \qquad
    \includegraphics[width=.4\textwidth]{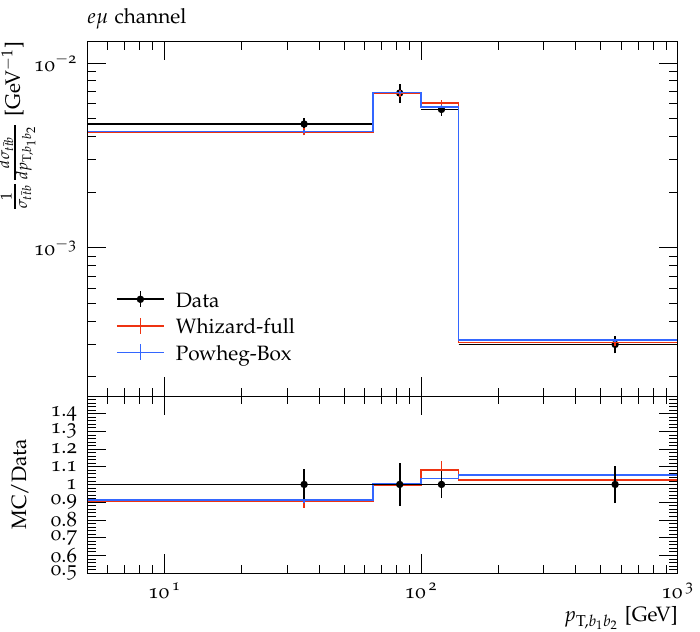}
  \end{center}  
  \caption{Left: comparison of different parton shower to LEP data for
    the differential distribution in the $4\to 5$ jet transition
    variable $y_{45}$, from~\cite{Herren:2022jej}. Right: comparison of
    the two NLO generators \textsc{POWHEG-Box} and \textsc{Whizard} on
    the matched $pp \to b\bar{b}e^-\mu^+\bar{\nu}_e\nu_\mu$ process at
    13 TeV to ATLAS data.}
  \label{fig:shower_matching}
\end{figure}
There are different strategies achieve this goal: veto emissions from
the parton shower into phase space regions better described by emissions from the hard matrix
elements (and potentially reweight by Sudakov non-emission
probabilities), or explicitly subtract from the fixed-order
calculations the parts described by the parton shower. An incarnation
of the first approach is the POWHEG
method~\cite{Nason:2004rx,Frixione:2007vw}), of the latter approach is
the MC@NLO
method~\cite{Frixione:2002ik,Torrielli:2010aw,Frixione:2010ra}. The
right panel in \figref{fig:shower_matching} shows an comparison of
different incarnations of the POWHEG matching algorithm to ATLAS data
for the process $pp \to b\bar{b}e^-\mu^+\bar{\nu}_e\nu_\mu$ at 13 TeV
LHC Run2. Other algorithms can be found
in~\cite{Jadach:2015mza,Nason:2021xke}. 

Since the mid-2010s, many efforts have been undertaken to combine NNLO
calculations with parton showers through
algorithms~\cite{Hoche:2014uhw} that have been 
carried also into MC frameworks and tools like MiNNLO$_{\rm
  PS}$~\cite{Monni:2019whf,Monni:2020nks} or 
GENEVA~\cite{Alioli:2015toa,Alioli:2012fc}.
With the advent of the first N3LO calculations (e.g. for Higgs
production in gluon function, cf. the section
in~\cite{Heinrich:2025ylf}), parton showers also need to be matched to
such cases where up to three emissions can come from the fixed order
calculation~\cite{Bertone:2022hig}. 

For many studies at LHC, it has turned out beneficial to use inclusive
multi-jet samples which contain a minimal number of selected jets
(zero, one, two), but are inclusive in events with all higher jet
multiciplities. In order to generate such samples, MC generators need
to merge samples with exclusive jet multiplicities at different orders
in the underlying fixed-order calculations (mostly LO and NLO). There
are several different merging algorithms for inclusive multi-jet
samples, again either discarding events where the shower would
populate already existing multiplicites, or re-using as many events as
possible and reweighting to get the normalization as close to the
experimental data (examples at LO
are~\cite{Catani:2001cc,Hoeche:2005vzu,Alwall:2007fs,Hamilton:2009ne},
while NLO applications are~\cite{Lavesson:2008ah,Frederix:2012ps,Kallweit:2015dum}).

Another important information that needs to be carried from the hard
process to the parton shower are decays of intermediate resonances
like the $W$, the $Z$, the Higgs boson and the top quark. Even when
the decays are included completely in full matrix elements (e.g. $2\to
6$ for top pairs or $2\to 4$ for hadronic $WW$), the MC generator
needs to provide a resonance history as close as possible to the
correct probabilities of the intermediate resonances: e.g. in $e^+e^-
\to jjjj$ at 240 GeV, roughly 79\% comes from $W$ pairs, 18\% from $Z$
pairs and 2\% is non-resonant QCD background. As the parton shower
does rearrange kinematics in order to implement certain recoil schemes
in its splittings or tries to bring partons close in phase space
before hadronization, it would shift invariant masses of combinations
of partons away from their original resonance masses. So the MC
generator needs to take care of this information. An example for this
is~\cite{Jezo:2016ujg}.

%%%%%

\subsection{QED showers and their matching, weak showers}
\label{sec:qedshower}

There are many collider observables, where a very precise simulation
of photon radiation and other QED effects are needed to match the
experimental precision: two paradigm examples are (i) the $W$ mass
measurements at hadron colliders like Tevatron and LHC, where the
transverse momentum distribution of the decay lepton is very sensitive
to final state QED showering, and (ii) the luminosity measurement at
$e^+e^-$ colliders relying on the most precise predictions of Bhabha
scattering, $e^+e^- \to e^+e^-$, as well as two-photon production,
$e^+e^- \to \gamma\gamma$. Many of these specialized QED shower
programs have been developed for the combined LEP1+2 program and the
flavor factories, BaBar and Belle~\cite{Jadach:1995nk,Jadach:1996is}. 

Also, many algorithms and implementations for QED showers originated
during the LEP time~\cite{CarloniCalame:2001ny}. Like for QCD, it is
important to properly combine these showers with fixed-order
calculations without double counting~\cite{Balossini:2006wc}. One of
the frameworks is Yennie-Frautschi-Suura (YFS) resummation based on
purely soft emissions~\cite{Yennie:1961ad}. This formalism allows both
to resum soft emissions to all orders in an inclusive way as well as
exclusive emissions of (soft) photons as a shower. This is the basis
for the implementations of many different LEP MC
generators~\cite{Skrzypek:1995wd,Jadach:1998gi,Jadach:1999tr,Jadach:2001uu,Jadach:2001mp,Jadach:2022mbe}
for 
two-particle final states, where the interference with hard emissions
in a coherent picture between initial- and final-state emissions is
possible~\cite{Jadach:2000ir}. More recently, the non-coherent
formalism has been automated for arbitrary
processes~\cite{Krauss:2022ajk}. A review of the corresponding tools
can be found here~\cite{Heinemeyer:2021rgq}. 
One of the more
ubiquitous tools for QED showering which is very often attached to the
simulations of multi-purpose generators, is
\textsc{Photos}~\cite{Barberio:1990ms,Barberio:1993qi,Davidson:2010ew}. 
Like for QCD, also for QED photon emissions from the shower need to be
properly matched to the fixed order hard process. Similar algorithms
as for QCD can be applied for QED, but there are also dedicated
algorithms that take the characterics of the emissions from the
analytic lepton PDFs (cf. \subsecref{sec:ele_pdf}) into
account~\cite{Kalinowski:2020lhp}. Examples for matching of NLO EW
corrections to QED showers for $e^+e^-$ colliders can be found
here~\cite{Kilian:2006cj,Robens:2008sa}. 

With the plans for a very high-energy hadron collider of the order of
70-100 TeV (FCC-hh) or lepton colliders at the range of 10 TeV
center-of-mass energy, also the inclusion of weak radiation has been
studied~\cite{Kleiss:2020rcg,Brooks:2021kji}. Weak splitting kernels can be included in the Sudakov
factors, and the showers can be interleaved with the QCD and QED
radiation. The DGLAP evolution then in principle connects all
components of the Standard Model (SM). An important feature is the
fact that weak splittings are chiral, so within weak showers partons
automatically become polarized.

%%%%%%%%%%%%%%%%%%%%%%%%%%%%%%%%%%%%%%%%%%%%%%%%%%%%%%%%%%%%%%%%%%%%%%%%%%%%%%%%
\section{Hadronization}
\label{sec:hadronization}

QCD becomes strongly interacting at low energy scales, and quarks
(with the exception of the top) and gluons form bound states, mesons,
baryons or more exotic objects (they ``hadronize'' or
``fragment''). While it is possible to calculate very inclusive
quantities analytically (cf. Sec. 7 in~\cite{Stagnitto:2025air}),
exclusive events need to be simulated by MC generators. As this is an
intrinsincially nonperturbative phenomenon, no miscroscopic
description from first principles exists and the fragmentation needs
to be modelled phenomenologically, with the parameters of these models
then being tuned to data. In \subsecref{sec:multipurpose}, we discuss
the two main existing fragmentation models, before we briefly comment
in Sec.~\subsecref{sec:hadrondec} on the simulation of hadronic decays and
QED radiation of hadrons. 

\subsection{Fragmentations of partons into hadrons in MC generators}
\label{sec:multipurpose}

After a first era of very crude models of hadronization, using
e.g. flux tubes as modelling of jet masses initiated by Feynman in the
early 1970s, the first serious hadronization model was the so called
independent fragmentation~\cite{Field:1976ve,Field:1977fa}, encoded in
the first hadronization program \textsc{Isajet}~\cite{Paige:1986vk} in
1979. This algorithm creates quark pairs from the vacuum to dress bare
quarks and uses a Gaussian distribution for the generation of
transverse momentum, $p_T$ , but suffers from several problem, especially Lorentz
covariance and infrared safety issues; also notoriously, the last open
quark needed special treatment. Independent fragmentation is still
used for determining total hadronic cross sections.

With the early 1980s, the two models which dominate fragmentation
simulations until today, were developed and put into codes:
fragmentation based on the Lund string model~\cite{Andersson:1983ia}
and cluster fragmentation
All MC generators today use variants of one of these two formalisms. 
The Lund string model became the basis of the
\textsc{Jetset}~\cite{Sjostrand:1984ic,Sjostrand:1986hx} event
generator which became fused into the
\textsc{Pythia}~\cite{Sjostrand:1985xi,Sjostrand:2006za} parton shower
simulation. All fragmentation models are based on the two fundamental
concepts of (1) local parton-hadron locality~\cite{Azimov:1984np}, assuming that fragmmentation
is process with low momentum transfer, such that energy momentum and
flavor quantum numbers of the produced hadrons follow closely those of
the underlyind partons; and (2) of a universal low-energy strong
coupling $\alpha_s$ used in all jets and
branchings~\cite{Marchesini:1983bm,Dokshitzer:1995zt}.

The Lund string fragmentation model is based on the first attempt to understand
strong interactions in terms of quantized strings, which can be
connected in the framework of QCD as the fact that QCD field lines at
low energies of a few $\Lambda_{\text{QCD}}$ and below become
compressed tubelike regions that correspond to the dynamics of
string-like objects. In that picture, the linear confinement potential
can be derived. Partons (quarks) that are close in phase space are
ordered and get connected by rubber-band like 
string objects. The masses of the generated mesons are roughly
proportional to the area that these strings trace in their time
evolution. Non-perturbative effects appear as Lund string breakings
which are simulated as tunneling processes, where in the breakup a new
quark-antiquark pair pops out of the vacuum. This fragementation
starts in the middle of the strings and spreads outward. The flavor
decomposition in fragmentation is roughly modelled as $u\bar{u} :
d\bar{d} : s\bar{s} : c\bar{c} \approx 1:1:0.3:10^{-11}$ while bottom
(and top) quarks are not generated in fragmentation. Though the Lund
string model aims at a universal description of data (e.g. $e^+e^-
\to \text{hadrons}$, it has a plethora of free parameters (roughly
$\mathcal{O}(1)$ per known hadron, and it has difficulties in
precisely describing baryon production (which tries to interpret
antiquark as diquark to group them with a quarks, modelled as string
junctions).

The second major hadronization formalism is cluster
fragmentation~\cite{Webber:1983if,Webber:1986mc,Marchesini:1987cf}
which became fused with the \textsc{Herwig} multi-purpose MC
generator~\cite{Corcella:2000bw}, and in a different incarnation provides
the hadronization module of the \textsc{Sherpa}
framework~\cite{Gleisberg:2008ta}. Cluster fragmentation makes use of
the fact that parton showers (at least their leading-logarithmic or
leading-color) part order partons in color space, such that also color
partners get generated close to each other in phase space, which
motivates taking the $N_c \to \infty$ limit where planar Feynman
diagrams dominate~\cite{tHooft:1973alw}. The clusters in this model
are a recursively defined model of high-mass resonances whose spectrum is
defined by the parton shower, so perturbation theory, where the
primary cluster is independent of the production mechanism. An
important parameter is the scale where the parton shower stops and
hadronization sets is. The original algorithm is pure kinematics,
i.e. phase space, no spin info, while now there are attempts to also
include spin effects.

In the recent years, there are several attempts to use machine
learning techniques in order to learn some of the core components of
fragmentation modelling algorithms from (LHC) data. This would replace
the fragmentation functions which are parameterized by a certain
functional form and fitted to data by numerical distributions that
come out of pattern recognition algorithms adapted to LHC data. This
is a computationally interesting path, but it is open whether that can
ever lead to a better conceptual understanding of non-perturbative
low-energy regime of strong interactions.

\subsection{Hadronic decays and final-state QED radiation}
\label{sec:hadrondec}

The main hadronization modelling creates a setup of hadronic
resonances (mesons, baryons and also several classes of more exotic
bound states like tetraquarks, pentaquarks, hadronic molecules
etc.). Except for the proton, none of these states is stable on a
macroscopic scale, so MC generators need to model the transitions of
these states and also their decays. Over a century, these decays have
been measured in dozens of larger and smaller collider and fixed
target experiments and have been catalogued by the Particle Data Group
(PDG),~\cite{ParticleDataGroup:2024cfk} (and older editions). All of
these decays of ca. 200 resonances in the PDG with thousands of decay
channels need to encoded in the decay modules of multi-purpose event
generators to guarantee a precise simulation of the hadronic part of
particle physics. Let us illustrate the complexity and variety with an
example of a hadronic decay chain:
\begin{equation}
  B^{\ast 0} \stackrel{(1)}{\to} \gamma B^0 \stackrel{(2)}{\hookrightarrow} \overline{B}^0
  \stackrel{(3)}{\hookrightarrow} e^-\overline{\nu}_e D^{\ast +}
  \stackrel{(4)}{\hookrightarrow} \pi^+ D^0 \stackrel{(5)}{\hookrightarrow}
  K^- \rho^+ \stackrel{(6)}{\hookrightarrow} \pi^+\pi^0 \stackrel{(7)}{\hookrightarrow} 
  e^+ e^- \gamma
\end{equation}
The first step, (1), is a radiative electromagnetic decay, step (2) is
a weak mixing of $B$ mesons, (3) is a weak decay, step (4) a strong
decay, (5) another weak decay with a broadly smeared $\rho$ resonance
peak, step (6) is a decay of the $\rho^+$ that has been produced
polarized and where angular correlations are crucial, while the final
step is a 3-particle ``Dalitz'' decay with a sharply peaked
$m_{ee}$. All of the charged particles in these cascade ($D^{\ast \pm}$,
$K^\pm$, $\rho^\pm, \pi^\pm$) undergo QED radiation,
cf. Sec.~\subsecref{sec:qedshower}. There is a plethora of different,
mostly, kinematic features like form factors, Dalitz correlations
(phase space correlations), resonance peaks (sometimes with deviations
from Breit-Wigner due to interference with QCD continuum), but also
effects like spin correlations, weak and strong mixing, etc.

Among the leptons, tau decays are quite important because they allow
information on the spin quantum numbers of their mother particles,
especially from their hadronic decays, so these decays need to be
simulated with great care, including spin correlations, QED radiation
and exact kinematics~\cite{Jadach:1993hs}.

Implementing all of these features is a major task for any generator,
and given the fact, that LHC in the runs up to now has discovered new
hadronic resonances and many more rare decay channels an equally
important task to maintain and update the hadronic databases of each
generator.

%%%%%%%%%%%%%%%%%%%%%%%%%%%%%%%%%%%%%%%%%%%%%%%%%%%%%%%%%%%%%%%%%%%%%%%%%%%%%%%%
\section{Conclusions}
\label{sec:conclusions}

In this Chapter, we have provided a pedagogical introduction into the
complexity of Monte Carlo event generators for collider applications
in particle physics. They are the main workhorses for the
data analysis and interpretation of all existing collider experiments
and are equally indispensable for planning future collider experiments
and determining their physics potential. Generally, they are two
classes of such tools: (i) multi-purpose MC event generators which aim
at covering all aspects; hence, they are very extensive and
complicated structures as they encode all of our knowledge about
particle physics; and (ii) dedicated MC tools that focus on a single
or a few specific aspects. The main parts of a multi-purpose generator
are the hard scattering process at lowest and higher fixed orders in
perturbation theory, the interface to simulating beam structure and
PDF for hadron and lepton colliders, the underlying events, the parton
showers, the hadronization and the decay of all unstable particles and
resonances. MC generators constitute a sizeale fraction of the
computing ressources of collider experiments (for ATLAS and CMS each
ca. 20\% for the current LHC runs): constant optimization and porting
to heterogeneous computing architectures is necessary to enhance their
speed and reduce their power consumption. Collider physics experiments
run over several decades, and analyses need to be revisited decays
after the end of the runtime of collider experiments; this leads to a
heavy burden in maintenance and reproducability which cannot lay on a
few or even a single scientist: MC generators are the prime examples
for collaborational efforts in theoretical particle physics at the
interplay between physics and computational science.

\noindent {\bf Acknowledgements}.
Over the years I am very much indebted for many fruitful discussions
on Monte Carlo event generators and the physics connected to them to
Guido Altarelli (\textdagger), Alexander Belyaev, Thomas Binoth
(\textdagger), Eduard Boos, Simon Bra{\ss}, Pia Bredt, Jonathan
Butterworth, Matteo Cacciari, Stefan Dittmaier, Silvia Ferrario
Ravasio, Rikkert Frederix, Stefano Frixione, Stefan Gieseke, Tao Han,
Stefan H\"oche, Staszek Jadach  (\textdagger), Toma\v{s} Je\v{z}o,
Stefan Kallweit, Alexander Karlberg, Wolfgang Kilian, Frank Krauss,
Jonas Lindert, Leif L\"onnblad, Fabio Maltoni, Olivier Mattelaer,
Kirill Melnikov, Pier Monni, Stephen Mrenna, Zoltan Nagy, Paolo Nason,
Thorsten Ohl, Davide Pagani, Mathieu Pellen, Fulvio Piccinini, Simon
Pl\"atzer, Tilman Plehn, Stefan Prestel, Alan Price, Alexander Pukhov,
Peter Richardson, Gavin Salam, Steffen Schumann, Frank Siegert,
Torbj\"orn Sj\"ostrand, Peter Skands, George Sterman, James Stirling
(\textdagger), Melissa van Beekveld, Andr\'{e} van Hameren, Bennie
Ward, Zbigniew W\k{a}s, Bryan Webber, Marius Wiesemann, Malgorzata
Worek, and Marco Zaro, and many, many more.

Funding is acknowledged from the DFG under the German
Excellence Strategy-EXC 2121 “Quantum Universe”-
390833306, and the National Science Centre (Poland) un-
der OPUS research project no. 2021/43/B/ST2/01778.
 Furthermore, we acknowledge support from the COMETA COST Action CA22130,
by EAJADE - Europe-America-Japan Accelerator Development and Exchange Programme
(101086276), and from the Joint Research Programme by the
International Center for Elementary Particle Physics (ICEPP), the
University of Tokyo. We thank the Galileo Galilei Institute for
Theoretical Physics for the hospitality and the INFN for partial
support during the completion of this work.

\clearpage
\bibliographystyle{Numbered-Style}
\bibliography{mcgen}

\end{document}